\documentclass[12pt,preprint]{aastex}

\begin{document}

\newcommand{\sy}{{Seyfert }}
\newcommand{\etal}{{ et al. }}
\newcommand{\beq}{\begin{equation}}
\newcommand{\eeq}{\end{equation}}
\newcommand{\msun}{M_{\odot }}

\title{Black hole mass and accretion rate of active
galactic nuclei with double-peaked broad emission lines}
\author{Xue-Bing Wu, F. K. Liu} 
\affil{Department of Astronomy, Peking University, Beijing 100871, China}

\email{wuxb@bac.pku.edu.cn; fkliu@bac.pku.edu.cn}

\begin{abstract}
Using an empirical relation between the broad 
line region size and optical continuum 
luminosity, we estimated the black hole mass and 
accretion rate for 135 AGNs with double-peaked broad emission lines in two 
samples, one from the SDSS and the other  from a survey of radio-loud 
broad emission
line AGNs. With black hole masses in a range from $3\times 10^7M_\odot$ 
to $5\times 10^9M_\odot$, these AGNs have the dimensionless accretion rates 
(Eddington 
ratios) between 0.001 ad 0.1 and the bolometric luminosity between 
$10^{43}erg/s$ and $10^{46}erg/s$, both being significantly larger
than those of several previously known low-luminosity ($L_{bol}<10^{43}erg/s$)
double-peaked AGNs. The optical--X-ray spectra indices, 
$\alpha_{OX}$, of these  high-luminosity 
double-peaked AGNs is between 1 and 1.9, systematically larger
than that of low-luminosity objects which is around 1. Modest correlations (with Spearman's
rank correlation coefficient of  0.60)
of the $\alpha_{OX}$ value with the Eddington ratio  and bolometric luminosity 
have been found, indicating
that double-peaked AGNs with higher Eddington ratio or higher luminosity
 tend to have larger
$\alpha_{OX}$ value. Based on these results we 
suggested that the accretion process in the central region of some 
high-luminosity
double-peaked emission line AGNs (especially those 
with Eddington ratio larger than 0.01) is probably 
different from that of low-luminosity
objects  where a well-known  ADAF-like accretion flow was thought to 
exist. It is 
likely that 
the accretion physics in some 
high-luminosity double-peaked AGNs is similar to that in normal Type 1 AGNs,
 which is also supported by the presence of  possible big blue bumps in 
the spectra of some
double-peaked AGNs with higher Eddington ratios.
We noticed that the prototype double-peaked emission line AGN,
Arp 102B, having black hole mass of $10^8M_\odot$ and dimensionless accretion
rate of 0.001, may be an ``intermediate'' object between the high and low luminosity
 double-peaked AGNs. 
In addition, we found an apparent strong anti-correlation
between the peak separation of double-peaked profile and Eddington ratio (with Spearman's
rank correlation coefficient of $-0.79$).  But such an anti-correlation is probably
induced by a strong correlation between the peak separation and emission
line widths and needs to be confirmed by future works. If it is real, 
it may provide us another clue to understand why
double-peaked broad emission lines were hardly found in luminous AGNs with Eddington ratio
larger than 0.1.

\end{abstract}


\section{Introduction}

A small number of AGNs display double-peaked broad emission line profiles in
their optical spectra (Eracleous \& Halpern 1994, 2003). These double-peaked
emission lines are usually attributed to the line emission from some parts of 
accretion disks at a distance of several hundreds to thousands gravitational
radius $R_g$ 
(defined as $GM_{BH}/c^2$ where $M_{BH}$ is the black hole mass) 
away from the central 
black hole (Chen, Halpern \& Fillippenko 1989; Chen \& Halpern 1989; 
Halpern 1990). The rotation of photoionized materials in the disk
naturally results in two (redshifted and blueshifted) peaks of emission 
lines. Therefore these double-peaked line profiles are often regarded 
as one of the evidences for the existence of accretion disks in AGNs.
Although the disk origin of double-peaked emission lines is mostly
favorite, 
some other  alternatives, such as two broad line regions
involving supermassive binary black holes (Begelman, Blandford \& Rees 1980),
a bipolar outflow (Zheng, Binette \& Sulentic 1990), and an anisotropic 
illuminated broad line region (Goad \& Wanders 1996), also exist in 
the literature. 

So far about 150 double-peaked broad emission line AGNs have been discovered. 
In a completed survey, Eracleous \& Halpern (1994, 2003) have found
26 double-peaked broad emission line objects from 106 radio-loud AGNs. The 
double-peaked profiles of 60\% of these 26 objects can be well fitted with the
Keplerian rotating accretion disk model. Recently Strateva et al. (2003) 
presented a new sample of 116 double-peaked Balmer line AGNs selected from
the Sloan Digital Sky Survey (SDSS). The profiles of most of these objects can be
well fitted with the disk emission model, showing that the double-peaked
 emission lines are probably produced in the disk part with inner radius 
of several
hundreds of $R_g$ and outer radius of several thousands of $R_g$.  Although 
76\% of the SDSS double-peaked objects are radio-quiet, they are 1.6 times more
likely to be radio sources than other AGNs. The luminosities of these SDSS 
double-peaked sources are medium ($\sim 10^{44}erg/s$) and only 12\% of them 
are classified as LINERs (Strateva et al. 2003).
On the other hand, double-peaked broad emission lines
have been also found in some nearby galaxies, including NGC 1097 (Storchi-Bergmann,
Baldwin \& Wilson 1993), M81 (Bower et al. 1996), NGC 4450 (Ho et al. 2000),
and NGC 4203 (Shields et al. 2000). These objects are all classified 
spectroscopically as Type 1 LINERs and have much lower nuclear luminosity
($<10^{43}erg/s$) and  smaller Eddington ratios ($<10^{-3}$)(Ho et al. 2000). 
Together with their relatively flat broad band continuum (Optical--X-ray 
spectral index
$\alpha_{OX}=0.9-1.1$, Ho \etal 2000), these properties suggests that
the accretion process in the nuclei of these low-luminosity 
objects may be in a form of 
advection-dominated accretion flow (ADAF, see Narayan,  Mahadevan \& Quataert 1998
for a review).  However, it is still not clear whether the high-luminosity 
 (with 
bolometric luminosity $L_{bol}>10^{43}erg/s$) AGNs with double-peaked 
broad emission lines have the same accretion physics
as the low-luminosity ones. Strateva et al. (2003) have shown that 
$\alpha_{OX}$ values for SDSS double-peaked AGNs are in the range of 1 to 2, with
an average value of 1.4, which is
significantly larger than that of low-luminosity double-peaked sources 
($\alpha_{OX}\sim 1$) but  similar to that of normal type 1 AGNs (Mushotzky \&
Wandel 1989).
This implies that the accretion physics may be different in the high-luminosity 
and low-luminosity double-peaked AGNs.

To understand better the nature of the double-peaked AGNs, we need to know some 
fundamental parameters of them, especially the black hole mass and the accretion rate.
Unfortunately, these parameters were known only for 4 low-luminosity sources 
(NGC 1097, M81, NGC4203, NGC 4450) and
3 high-luminosity sources (Arp 102B, Pictor A and 3C390.3) 
(see Wandel, Peterson \& Malkan (1999) for 3C390.3 and Ho et al. (2000) for 
others).
In this paper we  estimate the black hole mass and Eddington ratio for 135
double-peaked AGNs in section 2 and compare the different physical properties 
of high-luminosity and low-luminosity double-peaked AGNs in section 3. We 
give the summary and discussion of our results in Section 4.

\section{Black hole mass and accretion rate estimations}

Several methods have been adopted to estimate the central black hole masses 
of AGNs. With the dynamic methods, black hole masses of several nearby AGNs, including
a double peaked source M81 (NGC 3031), have been 
estimated (Kormendy \& Gebhardt 2001). Black hole masses of about 40 bright 
Seyfert 1 galaxies and nearby quasars have been derived with the  
reverberation mapping technique (Wandel, Peterson \& Malkan 1999; Ho 1999; Kaspi et al.
2000). Shortly after the confirmation of the applicability of a tight relation
between the black hole mass and the bulge velocity dispersion to AGNs 
(Ferrerase et al. 2001; Tremaine \etal 2002), a method of using such a relation to derive the
black hole masses of AGNs from the measured central velocity dispersion was
suggested and  applied to about 70 Seyfert
galaxies (Wu \& Han 2001a; Woo \& Urry 2002). However, these methods are 
applicable to only several double-peaked AGNs and there is  no estimate
of black hole mass for most double-peaked AGNs.

Reverberation mapping studies of 34 AGNs revealed a significant correlation 
between
the broad line region (BLR) size and the optical continuum luminosity 
(Kaspi et al. 2000), namely, 
\beq
R_{BLR}=32.9[\frac{L_{5100\AA}}{10^{44}erg/s}]^{0.7}~ \rm{lt-days} .
\eeq
Interestingly, a double-peaked broad line AGN, 3C 390.3 (with 
$R_{BLR}=22.9~ \rm{light days}$ and $L_{5100\AA}=6.4\times10^{43}erg/s$), 
is one of these
AGNs and its observational data is well consistent with such an empirical
relation. This may imply that we can use the R-L relation to estimate the
BLR size from the optical continuum luminosity of high-luminosity 
double-peaked AGNs.  On the other hand, the width (FWHM) of Balmer lines can be
used to estimate the velocity dispersion of the BLR. Therefore, The virial
mass of the central black hole can be estimated with
\beq
M_{BH}=1.464\times10^5(\frac{R_{BLR}}{\rm{lt-days}})(\frac{V_{FWHM}}
{10^3km/s})^2 \msun ,
\eeq
where $V_{FWHM}$ is the FWHM value of Balmer emission lines.

In the reverberation mapping studies, both $R_{BLR}$ and $V_{FWHM}$
are measured mostly from the broad H$_\beta$ emission line. Previous
study has shown that there is a relation between the widths of H$_\alpha$
and H$_\beta$ emission lines, namely $V_{FWHM}({\rm{H_\alpha}})=0.873V_{FWHM}
({\rm{H_\beta}})$ (Eracleous \& Halpern 1994). Kaspi \etal (2000) 
also derived a linear relation 
$R_{BLR}({\rm{H_\alpha}})=1.19R_{BLR}({\rm{H_\beta}})+13$, which is
equivalent to $V_{FWHM}({\rm{H_\alpha}})=0.91V_{FWHM}({\rm{H_\beta}})$,
consistent with the previous result. For the double-peaked AGNs in different 
samples, usually we have the measurement values of
their broad H$_\alpha$ emission line widths. By subtracting the starlight
contribution and considering the Galactic extinction, we can also derive the
optical continuum luminosity from their apparent magnitudes. Therefore  the 
black hole masses of double-peaked AGNs can be obtained with Eqs. (1) and (2).
Assuming a relation between the bolometric luminosity and the optical continuum
luminosity, $L_{bol}\simeq 9L_{5100\AA}$ (Kaspi \etal 2000), we can estimate
their Eddington ratios (defined as $\dot{m}=L_{bol}/L_{Edd}$ where $L_{Edd}$ is the
Eddington luminosity given by $L_{Edd}=1.26\times10^{46}erg/s (M_{BH}
/10^8\msun)$), which measure the accretion rates of AGNs in Eddington unit.
We noticed that the above formula for the conversion between the bolometric 
luminosity and optical nuclear continuum luminosity was obtained mainly for bright 
type 1 AGNs, which may not be applicable to all double-peaked AGNs. However, 
for high-luminosity double-peaked AGNs, the overall SED may not be too 
different from normal AGNs. Strateva et al. (2003) have  shown that the 
average value of broad band spectral index $\alpha_{OX}$ for the SDSS 
double-peaked AGNs is similar to normal type 1 AGNs. 
Only 12\% of these SDSS AGNs can be classified as LINERs and most
of them may be more similar as luminous broad line AGNs. 
Therefore,
if the starlight contribution is subtracted, the average SED of 
high-luminosity double-peaked AGNs may be similar to
that of type 1 AGNs and the difference in the conversion factor between the 
bolometric luminosity and nuclear optical continuum luminosity 
would not significantly change our main results.

Now we apply the method  to derive the black hole masses and
accretion rates of double-peaked AGNs in the SDSS. Strateva et al. (2003) have
given the magnitudes (Galactic extinction corrected), redshifts and FWHM values of
the double-peaked H$_\alpha$ line for 109 SDSS AGNs (7 objects have no FWHM
values among 116 AGNs in their table 3). From the $g$ and $r$ magnitudes of these AGNs
(available in Tables 1 and 2 in Strateva et al. (2003)), we can
estimate the rest frame flux at 5100$\AA$ by the following formula:
\beq
f_{5100\AA}(Jy)=3631\times10^{-0.4g}[\frac{4770}{5100(1+z)}]^{-\frac{g-r}
{2.5lg(6131/4770)}} .
\eeq
Because the starlight fraction values are not available for the SDSS double-peaked
AGNs, we assumed a typical value of starlight fraction of double-peaked AGNs as 0.33 
(Eracleous \& Halpern 2003) to estimate the luminosity of nuclear optical continuum at rest 
frame 5100$\AA$ (throughout the paper we adopted the Hubble constant as 
75$~{\rm{km/s/Mpc}}$ and the deceleration parameter
as 0.5). Using the conversion formula between the FWHM values of H$_\alpha$
and H$_\beta$ lines,$V_{FWHM}({\rm{H_\alpha}})=0.873V_{FWHM}
({\rm{H_\beta}})$, we can estimate the black hole masses of the double-peaked
SDSS AGNs with Eqs (1) and (2). The Eddington ratios of these AGNs can be also
derived if we adopted $L_{bol}\simeq 9L_{5100\AA}$ (Kaspi \etal 2000). 
The results and the related data of SDSS double-peaked AGNs are summarized in
Table 1. We can see that the black hole masses of these double-peaked AGNs
are in the range from $4\times10^7
\msun$ to $5\times10^9\msun$, and their Eddington ratios are between 0.002 and 0.2.
The average black hole mass of these 109 SDSS double-peaked AGNs is $10^{8.74}M_\odot$
and the average Eddington ratio is $10^{-1.82}$. The histograms of their black hole
masses and Eddington ratios  are shown in the upper panels of Fig. 1.

The same method is also applied to 26 double-peaked AGNs found from a survey of
 radio-loud emission line 
AGNs (Eracleous \& Halpern 1994, 2003). Assuming a power-law dependence of 
continuum flux on the frequency, $f_\nu\propto \nu^{-\alpha}$, the rest-frame
flux at 5100$\AA$ can be estimated from the V-band magnitude $m_V$ and Galactic
extinction $A_V$ (taken from Schlegel, Finkbeiner \& Davis 1998) by the 
following formula:
\beq
f_{5100\AA}(erg~ cm^{-2}s^{-1}\AA^{-1})=3.75\times10^{-9}[\frac{5500}{5100(1+z)}]
^{2-\alpha} 10^{-0.4(m_V-A_V)}.
\eeq
Using the values of V-band magnitude, redshift, starlight fraction of these
26 double-peaked AGNs given in Eracleous \& Halpern (1994, 2003)(except that 
we set
the starlight fraction as 0.9 rather than 1 for MS 0450.3-1817 and Arp 102B 
in order to estimate the nuclear continuum luminosity), and assuming
the spectral index $\alpha$ as about 0.5, we can estimate the nuclear rest-frame
continuum luminosity at 5100$\AA$. With the FWHM values of broad
H$_\alpha$ line of these AGNs given also in Eracleous \& Halpern (1994, 2003) and
adopting the conversion formula between the FWHM values of H$_\alpha$
and H$_\beta$ lines, we can estimate the black hole masses and Eddington
ratios of these 26 double-peaked AGNs. The results and the related data
are summarized in Table 2.
The black hole masses are from $3\times10^7
\msun$ to $3\times10^9\msun$ and the Eddington ratios are between 0.001 and 0.08,
The average black hole mass of these 26 double-peaked AGNs is $10^{8.78}M_\odot$
and the average Eddington ratio is $10^{-2.01}$, both being similar to those obtained 
for the double-peaked SDSS AGNs. The histograms of the black hole
masses and Eddington ratios of these 26 radio-loud double-peaked AGNs
 are shown in the lower panels of Fig. 1.

The black hole masses and Eddington ratios of 3 of the radio-loud, 
double-peaked AGNs have been estimated previously. For 3C 390.3, 
 the black hole mass of 3.7$\times10^8\msun$ and Eddington 
ratio of 0.012 were derived from the reverberation
mapping study (Wandel \etal 1998; Kaspi et al. 2000). Our estimation gives the values of 
7.3$\times10^8\msun$ and 0.007 respectively. For Arp 102B and Pictor A, Ho
et al. (2000) gave their black hole masses of 2.2$\times10^8\msun$ and
6.2$\times10^8\msun$, and Eddington ratios of 0.001 and 0.015 respectively.
Our estimations give the black hole mass of Arp 102B as 1.4$\times10^8\msun$
and that of Pictor A as 5.9$\times10^8\msun$, the Eddington ratio of Arp 102B
as 0.0014 and that of Pictor A as 0.002. Most of our estimated values are consistent
 with previous results within a factor of a few.

\section{Comparison of high-luminosity and low-luminosity AGNs with 
double-peaked broad emission lines}

Double-peaked broad emission line objects
have been found in some nearby galaxies, including NGC 1097 (Storchi-Bergmann,
Baldwin \& Wilson 1993), M81 (Bower et al. 1996), NGC 4203 (Shields et al. 2000)
and NGC 4450 (Ho et al. 2000). These objects are all classified 
spectroscopically as Type 1 LINERs and have very small nuclear luminosity
($<10^{43}erg/s$), very low Eddington ratios ($<10^{-3}$) and relatively flat 
broad band continuum ($\alpha_{OX}=0.9-1.1$) (Ho et al. 2000). 
However, the high-luminosity AGNs with double-peaked broad emission lines seem
to have some different properties from their low-luminosity counterparts.
Eracleous \& Halpern (1994, 2003) indicated that the X-ray luminosities of
most double-peaked objects in their radio-loud sample are larger than $10^{43}erg/s$,
implying that the nuclear luminosities of these objects are at least two orders
higher than
the low-luminosity double-peaked objects. Strateva et al. (2003) also noticed
that  most double-peaked SDSS AGNs have optical luminosities of 
a few times $10^{44}erg/s$, similar to the average value of all SDSS AGNs. They
found that 
only 12\% of these double-peaked AGNs are classfied as LINERs. The avearge value of
$\alpha_{OX}$  of these objects is 1.4, which is the same as that found for normal
type 1 AGNs (Mushotzky \& Wandel 1989) and normal SDSS AGNs (Anderson
\etal 2003), but significantly larger than that of low-luminosity double-peaked
AGNs. These differences between the high-luminosity and
low-luminosity double-peaked AGNs may not be purely attributed to the 
selection effects. 

After deriving the black hole masses and Eddington ratios for the high-luminosity
double-peaked AGNs in two samples, we are able to compare these fundamental parameters
with those of low-luminosity double-peaked objects, which were summarized in Ho
\etal (2000). In Fig. 2 we show the values of black hole masses and Eddington ratios
of these double-peaked AGNs. Clearly we see that all the 4 low-luminosity objects
have black hole masses smaller than $10^8M_\odot$ and Eddington ratios lower
than $10^{-3}$, while most of high-luminosity objects have black hole masses larger 
than $10^8M_\odot$ and all of them have Eddington ratios higher than $10^{-3}$. 
The average value of black hole masses of high-luminosity objects is about one
order larger than that of the low-luminosity objects, and the average value of
Eddington ratios of the former objects is about two orders higher than the later 
ones.
We also noticed that the
prototype object of double-peaked AGNs, Arp 102B, with black hole mass of 
 $10^{8.1}M_\odot$ and Eddington ratio of $10^{-2.8}$, locates right between
the low-luminosity and high-luminosity objects in Fig. 2, meaning that it may be
an ``intermediate'' object. In Fig 3 we show the distributions of
the bolometric luminosities of these double-peaked AGNs. 
We can see that the low-luminosity double-peaked AGNs have both lower bolometric
luminosities and lower Eddington ratios, in contrary to the high-luminosity objects.
The average value of bolometric luminosities of high-luminosity objects is more than
two orders higher than that of low-luminosity objects.
Again, Arp 102B, with the bolometric luminosity of $2.7\times10^{43}erg/s$, 
 locates right between these low-luminosity
and high-luminosity double-peaked AGNs in Fig. 3.

As we have mentioned, there are also differences in the optical--X-ray $\alpha_{OX}$
spectral indices
of high and low luminosity double-peaked AGNs. This may  reflect the difference
in their broad band continuum shapes, which are closely related to the radiation processes in
the center of these AGNs. The $\alpha_{OX}$ values are available for 47 SDSS 
double-peaked AGNs (Strateva et al. 2003)
 and 4 low-luminosity ones (Ho \etal 2000). Using our estimated nuclear luminosity at
5100$\AA$ (see section 2) and the X-ray luminosity data (mostly in the 0.1-2.4keV band)
for 26 double-peaked AGNs in the radio-loud AGN
sample (Eracleous \& Halpern 2003), we can derive the luminosity values at 2500$\AA$ and 2keV
and then estimate the   $\alpha_{OX}$ values for these  objects assuming a typical optical
continuum spectral index of 0.5 and a typical X-ray photon index of 2. The derived $\alpha_{OX}$
values, in a range from 1 to 1.7 and with an average value of 1.2,  are also listed in Table 2.
Evidently, the $\alpha_{OX}$ values of double-peaked AGNs in two high-luminosity samples are
larger than those of low-luminosity objects. In Fig. 4 we show the dependence of 
$\alpha_{OX}$ value on the Eddington ratio for all these 77 double-peaked AGNs. 
A Spearman's rank test gives the correlation coefficient of  0.60 and a chance
 probability of
 $3.5\times10^{-8}$, indicating a modest correlation between the $\alpha_{OX}$ values and Eddington
ratios.
From Fig. 4 we can clearly observe
a trend that the objects with lower Eddington ratios seem to have smaller $\alpha_{OX}$ values. When
the Eddington ratio of an  object is higher than 0.01, it usually has 
$\alpha_{OX}$ value larger than 1. We noticed that one of the SDSS double-peaked AGNs, SDSS J1710+6521, 
with the largest Eddington 
ratio of $10^{-0.9}$ in the SDSS sample, also has the largest $\alpha_{OX}$ value of 1.9. 
 However, we noticed that both the $\alpha_{OX}$ and Eddington ratio depend
on the optical continuum luminosity. If we keep the continuum luminosity at $5100\AA$ fixed, the partial correlation coefficient between the  $\alpha_{OX}$ and Eddington ratio is only 0.24. Therefore, the modest correlation between
the $\alpha_{OX}$ and Eddington ratio is probably induced by the correlation
between the $\alpha_{OX}$ and optical continuum luminosity (or bolometric
luminosity), which showes a
Spearman's rank corrleation coefficient of 0.60 and a chance probability
of $3.2\times 10^{-8}$ for 77 double peaked AGNs.

The differences in the Eddington ratios and $\alpha_{OX}$ values of these double-peaked AGNs with 
high and low luminosities may be also related to the presence or absence of a big blue bump in the
optical/UV continuum. Such a big blue bump is usually regarded as the signature of Type 1 AGNs 
(Sun \& Malkan 1989), and
most probably originates from the thermal radiation of an optically thick, geometrically thin
 accretion disk (Shakura \& Sunyaev 1973). In
Fig. 5 we show the sample spectra of four objects in the SDSS sample of double-peaked AGNs\footnote
{Spectra kindly available from Dr. I. V. Strateva at http://astro.Princeton.EDU/~iskra/original.spectra.tgz}. Two of them,
SDSS J1710+6521 and SDSS J1424+5953, have the largest Eddington ratios in the sample, while the other
two AGNs, SDSS J0817+3435 and SDSS J0759+3528, have the smallest Eddington ratios. From their 
spectra we can clearly see that two AGNs with larger Eddington ratios have relatively 
stronger radiation in the blue-ward of $5000\AA$ than other two with smaller Eddington ratios. 
Such a difference in the spectra shape can be also clearly seen from their $u-r$ colors. The 
 $u-r$ values are 0.10 and 0.13 for SDSS J1710+6521 and SDSS J1424+5953, while those are
2.53 and 0.92 for SDSS J0817+3435 and SDSS J0759+3528 respectively.
Therefore it is clear that a big blue bump is probably present in some high-luminosity 
double-peaked AGNs with larger Eddington ratio. The properties of these high-luminosity 
objects may be  similar to type 1 AGNs in many aspects. 
On the contrary, the low-luminosity double-peaked AGNs,
 most of which have Eddington ratio lower than $10^{-3}$, probably be lack of such a big 
blue bump in their LINER-like spectra (Ho \etal 2000).  

From our study we see the 
there is only one object
with bolometric luminosity larger than $10^{46}erg/s$ and with Eddington ratio larger than
0.1. It seems that we can hardly observe such an object in more luminous AGNs.
In some previous studies it has been suggested the reason for that is probably due to
the face-on orientation (Corbin 1997) or the broad emission lines arising
from an accretion-disk wind rather than from the disk itself (Murray \& Chiang 1997;
Eracleous \& Halpern 2003)
for majority of luminous AGNs. Here we propose another explanation based on
the investigation of the dependence of the peak separation
 of the double-peaked emission line profile on the Eddington ratio.
Strateva et al. (2003) have given the red peak and blue peak positions, $\lambda_{red}$
and $\lambda_{blue}$, for 116 double-peaked SDSS AGNs in their Table 3. The Eddington
ratios of 109 of these objects (other 7 objects have no published FWHM values) have been 
estimated by us in Section 2. In Fig. 6 we show the relation between the peak separation,
 $\Delta\lambda=\lambda_{blue}-\lambda_{red}$  (their values are listed in Table 1), and the Eddington ratio for these 109 
double-peaked SDSS AGNs. 
Clearly we see that as the 
Eddington ratio increases the peak separation decreases. A Spearman's rank test gives
a correlation coefficient of -0.79 and a chance probability of $5.2\times 10^{-24}$, 
suggesting a strong apparent anti-correlation between the peak separation and Eddington ratio.
If the Eddington ratio of a double-peaked AGN is higher than 0.1, its
peak separation will be probably smaller than 1000km/s (or 22\AA). The double peaks in 
the emission
line profile of this object will be very difficult to be detected. 

However, we noticed that there is a  strong correlation between the line peak 
separation and line width (FWHM) (with a Spearman's rank correlation coefficient of 0.84). This may natually lead to an anti-correlation between the line
peak separation and Eddington ratio because the black hole mass is proportional
to the square of FWHM. Keeping the H$_\beta$ line FWHM fixed, we found the partial correlation coefficient of line peak separation with the Eddington ratio is
only 0.05.  Therefore, we need to be cautious about such an apparent anti-correlation if we calculate the black hole mass using the emission line width. On
the other hand,  many studies have shown that narrow line
Seyfert 1 galaxies may have relatively smaller black hole masses and higher
accretion rates (Boller, Brandt \& Fink 1996; Mathur 2000;
Puchnarewicz et al. 2001), which seems to support 
the anti-correlation between the line width the Eddington ratio.
Future works, especially those on estimating the black hole mass without using the
emission line width, are needed to confirm such an anti-correlation.
If it  is 
real, such an anti-correlation 
may provide us a  clue 
to explain why the double-peaked broad emission line profiles were mostly
found in AGNs with low Eddington ratios.

\section{Summary and Discussion}

We have derived the black hole masses and Eddington ratios for 135 double-peaked broad line
AGNs in two samples( Strateva et al. 2003); Eracleous \& Halpern 1994, 2003).
 These estimations enable us to compare their properties with those
of several low-luminosity double-peaked AGNs known previously (Ho \etal. 2000). 
We found that these 135 double-peaked AGNs have black hole
masses from $3\times 10^7M_\odot$ 
to $5\times 10^9M_\odot$, dimensionless accretion rates (Eddington ratios)
from 0.001 to 0.1.  and bolometric luminosities from 
$10^{43}erg/s$ to $10^{46}erg/s$.  These values are significantly larger
than those of several  low-luminosity
AGNs with double-peaked broad emission lines. The optical--X-ray $\alpha_{OX}$
spectral indices of these double-peaked AGNs are from 1 to 1.9, with an average value of
1.4 and 1.2 respectively in two samples, being 
systematically larger 
than those of low-luminosity double-peaked AGNs. We have  found a modest correlation 
between the $\alpha_{OX}$ value and Eddington ratio for
 double-peaked AGNs and have shown that the double-peaked AGNs with higher 
Eddington ratios  (or higher luminosity) tend to have larger $\alpha_{OX}$ values. In addition, we 
demonstrated the differences in the continuum shape for double-peaked AGNs with
higher and lower Eddington ratios and found that the objects with larger Eddington
ratios probably display a big blue bump in their spectra, similar to many
normal type 1 AGNs. However,  an apparent anti-correlation was found between the peak separation and 
Eddington ratio. If such an anti-correlation could be 
confirmed by future works, it 
may help us to explain why the double-peaked profiles are hardly observed in more 
luminous AGNs with Eddington ratio higher than 0.1.

Our results suggested that high-luminosity double-peaked AGNs probably have some different
properties from low-luminosity ones. This is supported by the smaller fraction (~12\%) 
of LINER-like objects in the SDSS double-peaked AGNs (Strateva et al. 2003), but is
different from the previous suspicion that 
the
double-peaked broad line AGNs are  LINER-like objects with low-luminosity and 
low accretion rate (Eracleous \& Halpern 1994; Ho \etal 2000). Now we see that the
double-peaked
broad emission lines can also be found in AGNs with bolometric luminosity up to $10^{46}
erg/s$, with black hole mass up to $10^9M_\odot$ and with Eddington ratio up to
0.1. Some high luminosity double-peaked AGNs, especially those with Eddington ratio 
larger than 0.01, may share many similar properties, such as
a big blue bump and  a large $\alpha_{OX}$ value, as normal type 1 AGNs.
Our results support that the high-luminosity double-peaked AGNs (with Eddington ratio
larger than 0.01) may have different accretion disk structure with low-luminosity ones.
This can be clearly see from the relation of their  $\alpha_{OX}$ values with Eddington
ratios (see Fig. 4). Most probably, an ADAF-like accretion flow exists in
the low-luminosity double-peaked AGNs with Eddington ratio smaller than 0.01 while an
optically thick, geometrically think accretion disk exists in the high-luminosity 
double-peaked AGNs with Eddington ratios larger than 0.01. Such a critical Eddington
ratio or dimensionless accretion rate (around 0.01)  has been also mentioned in 
some theoretical studies on the transition between the hot ADAF and the cold disk 
models(Esin, McClintock \& Narayan
1997; Narayan \etal 1998; Rozanska \& Czerny 2000). The different radiation processes
in an ADAF and a cold disk may naturally account for the observed lower $\alpha_{OX}$ values 
of low-luminosity double-peaked AGNs and the higher values of high-luminosity ones. The absence
of a big blue bump in the low-luminosity double-peaked AGNs and the presence of that in
some high-luminosity double-peaked AGNs are also consistent with the different accretion process
in these objects. However, more detailed theoretical modelings of the accretion disk structure
and the broad emission line region of double-peaked AGNs are still needed to understand our  
derived relations of  the  $\alpha_{OX}$ value and the peak
separation with Eddington ratio.

Finally we would like to mention that there are substantial uncertainties in estimating the 
black hole masses and 
Eddington ratios of double-peaked AGNs. 
Although we believe that considering these uncertainties will not alter the main results 
in this paper,
we should be cautious in using the estimated value of black hole mass and Eddington ratio
for a specific object. 
Firstly, the poor understandings of the geometry and 
dynamics of the broad line region may cause substantial errors in black hole mass estimation
(Krolik 2001). The inclination of the broad line region, which may not be randomly distributed
for double-peaked AGNs, may lead to an uncertainty in black hole mass 
up to a factor of a few (McLure \& Dunlop 2001; Wu \& Han 2001b). Secondly, the variations of the
broad emission line profiles are common in double-peaked AGNs (Zheng, Veilleux \& Grandi 1991; 
Newman \etal 1997; Storchi-Bergmann \etal 2003). This may also affect the accuracy
of black hole mass
estimation using the Virial law. Thirdly, the starlight fraction for double-peaked AGNs
varies a lot from 0.1 to 1 (Eracleous \& Halpern 1994). A careful subtractions of the starlight
from the observed luminosity must be performed in order to accurately estimate the nuclear
continuum luminosity, which is important in deriving the broad line region size and the
bolometric luminosity for a double-peaked AGN. Finally,  in this paper the  bolometric
luminosity of double-peaked AGNs is simply derived from the nuclear continuum luminosity 
at $5100\AA$. More accurate
estimations may be done by integrating the observed flux points in a SED from radio to 
X-ray or by fitting the average SED for double-peaked AGNs to the available flux points
(Woo \& Urry 2001). Future efforts in diminishing these uncertainties will be
undoubtedly much helpful to understand the physics of double-peaked broad line AGNs.  

\acknowledgments
 
 We are grateful to the anonymous referee for helpful suggestions which improve
the presentation of our paper. The work is supported by
the National Key Project on Fundamental Researches (TG 1999075403),
the National Natural Science Foundation (No. 10173001 and N0. 10203001) in China .

\newpage
\begin{figure}
\plotone{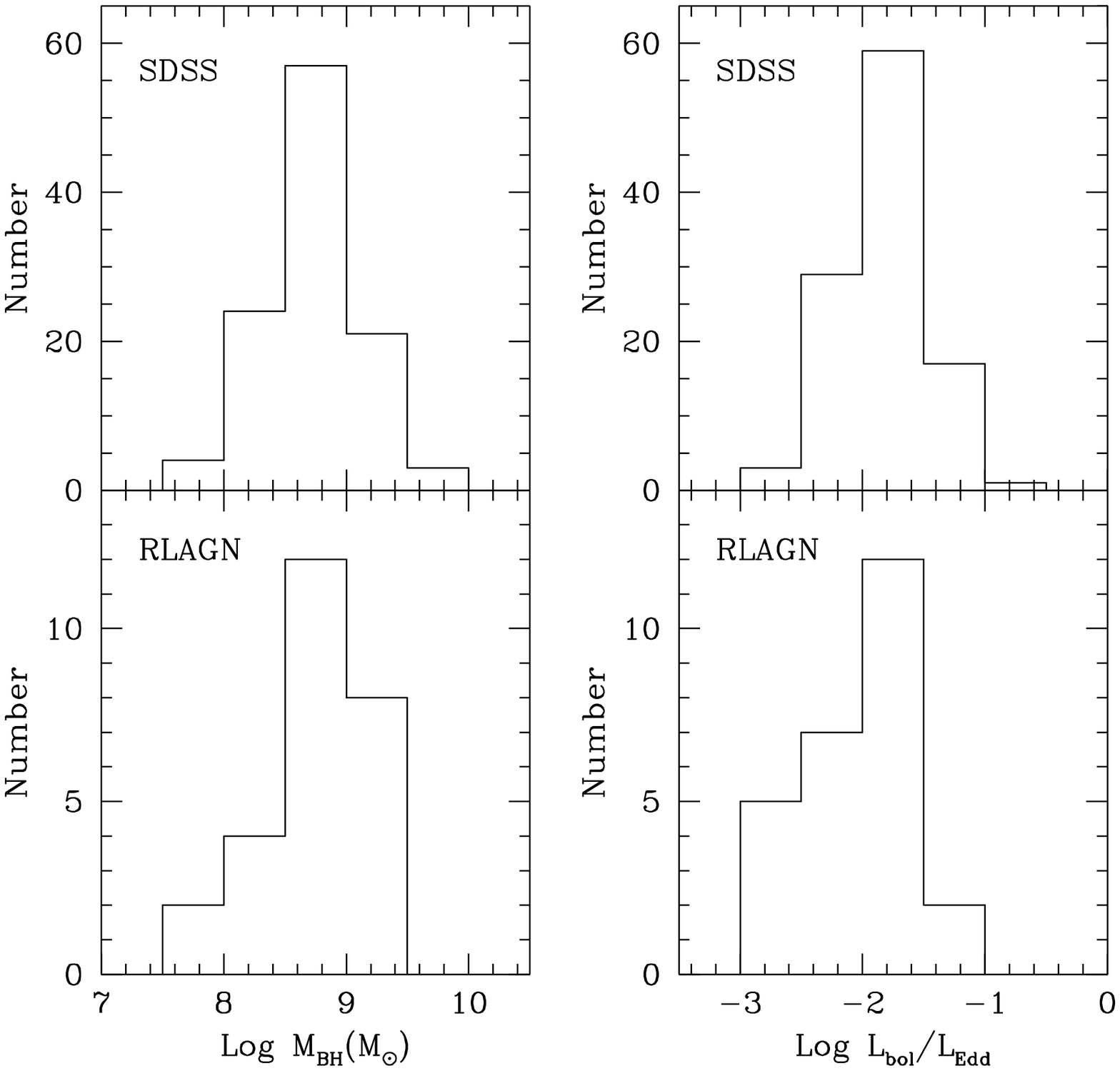}
 \caption{The histograms of the black hole masses and Eddington ratios of 
double-peaked AGNs in the SDSS (Strateva et al. 2003) and radio-loud AGN 
(Eracleous \& Halpern 1994, 2003) samples.
} 
\end{figure}

\begin{figure}
\plotone{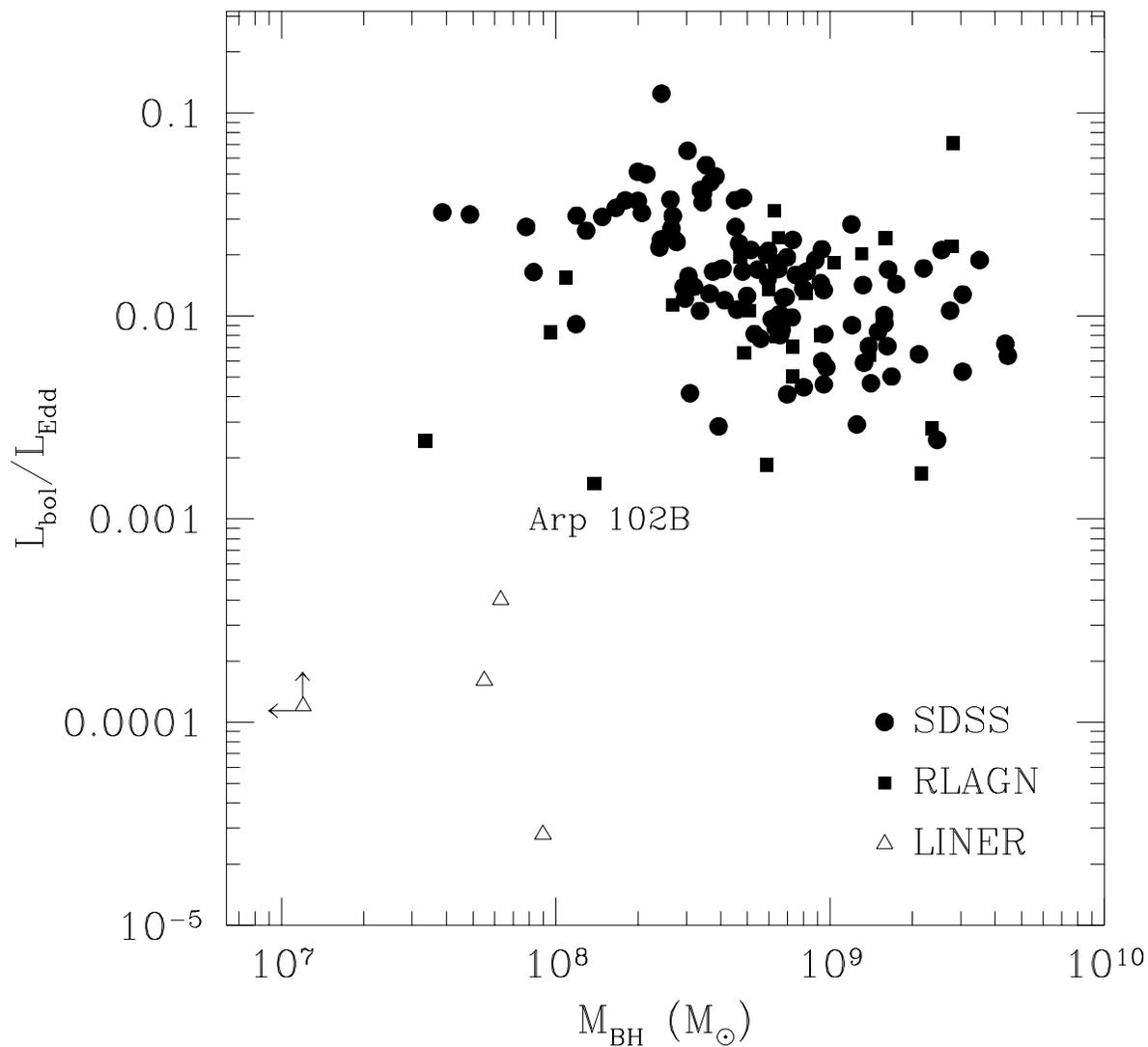}
 \caption{Black hole mass Eddington ratio distributions in  three 
different samples of double-peaked AGNs. The solid circles and squares 
represent the sources in the SDSS (Strateva et al. 2003) 
and the radio-loud AGN survey (Eracleous \& Halpern 1994, 2003) respectively.
The open triangles represent four low-luminosity sources spectroscopically 
identfied as LINERs (Ho \etal 2000).}
\end{figure}

\begin{figure}
\plotone{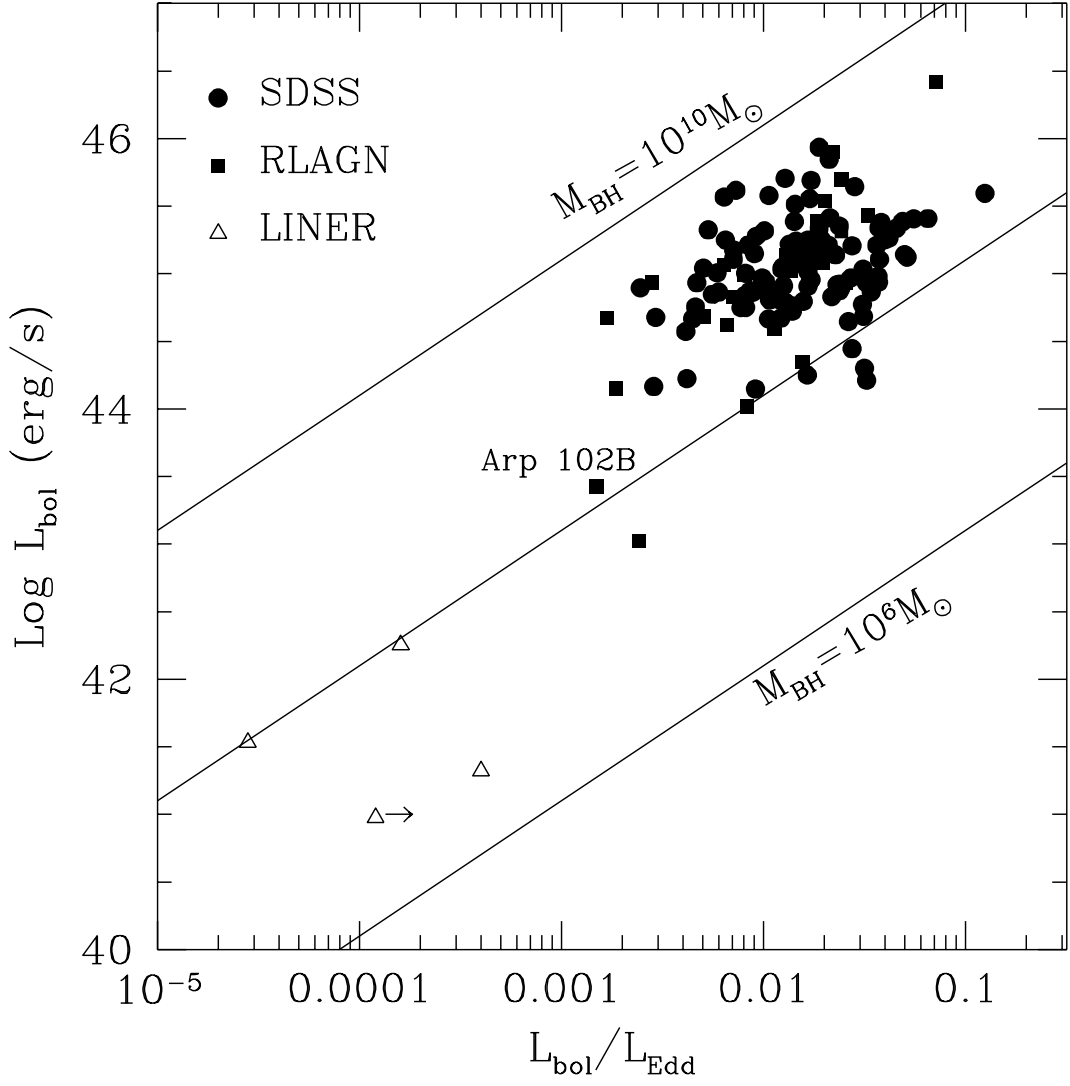}
 \caption{Relation between bolometric luminosity and Eddington ratios for 
three 
different samples of double-peaked AGNs. The symbols have the same meanings
as in Fig. 2. The solid lines represent the relations for the cases with
black hole mass of $10^6M_\odot$, $10^8M_\odot$ and $10^{10}M_\odot$ respectively.}
\end{figure}

\begin{figure}
\plotone{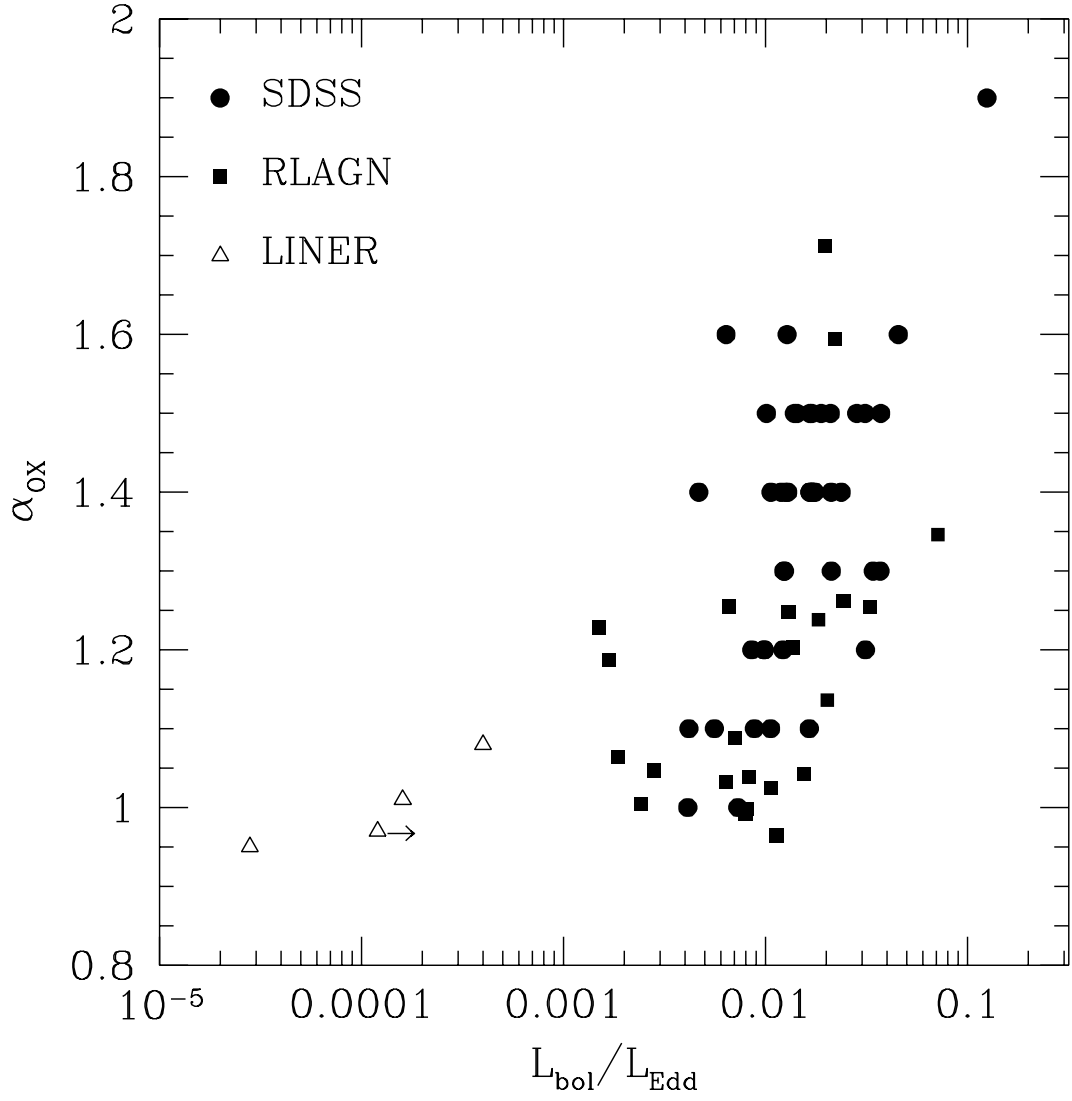}
 \caption{Relation between $\alpha_{OX}$ and Eddington ratios for three 
different samples of double-peaked AGNs. The symbols have the same meanings
as in Fig. 2.}
\end{figure}

\begin{figure}
\plotone{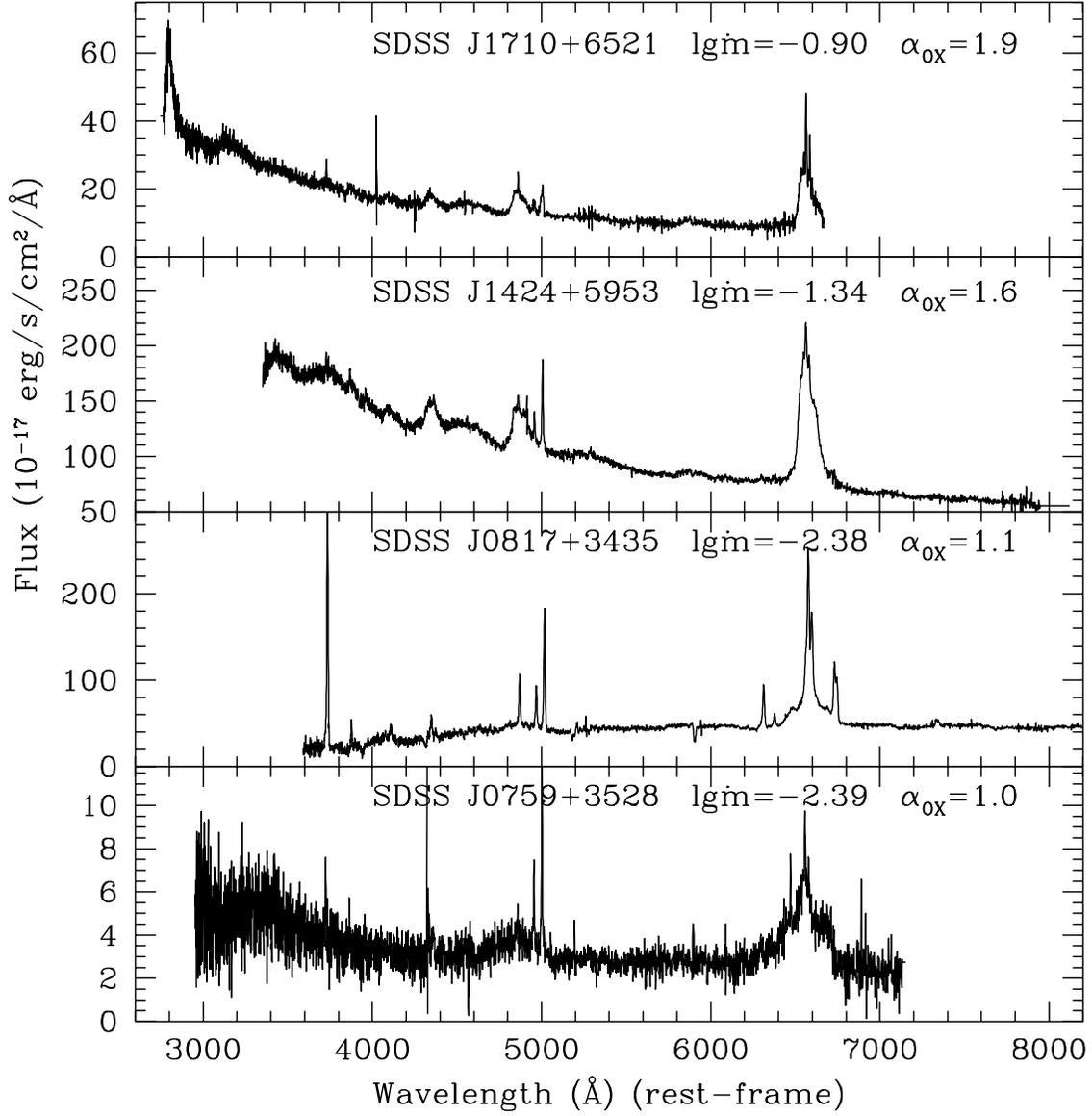}
 \caption{Comparison of sample spectra of four double-peaked SDSS AGNs. Two AGNs in 
the two upper panels have largest Eddington ratios ($\dot{m}=L_{bol}/L_{Edd}$) 
while two in the two lower panels
have smallest Eddington ratios in the SDSS sample.}
\end{figure}

\begin{figure}
\plotone{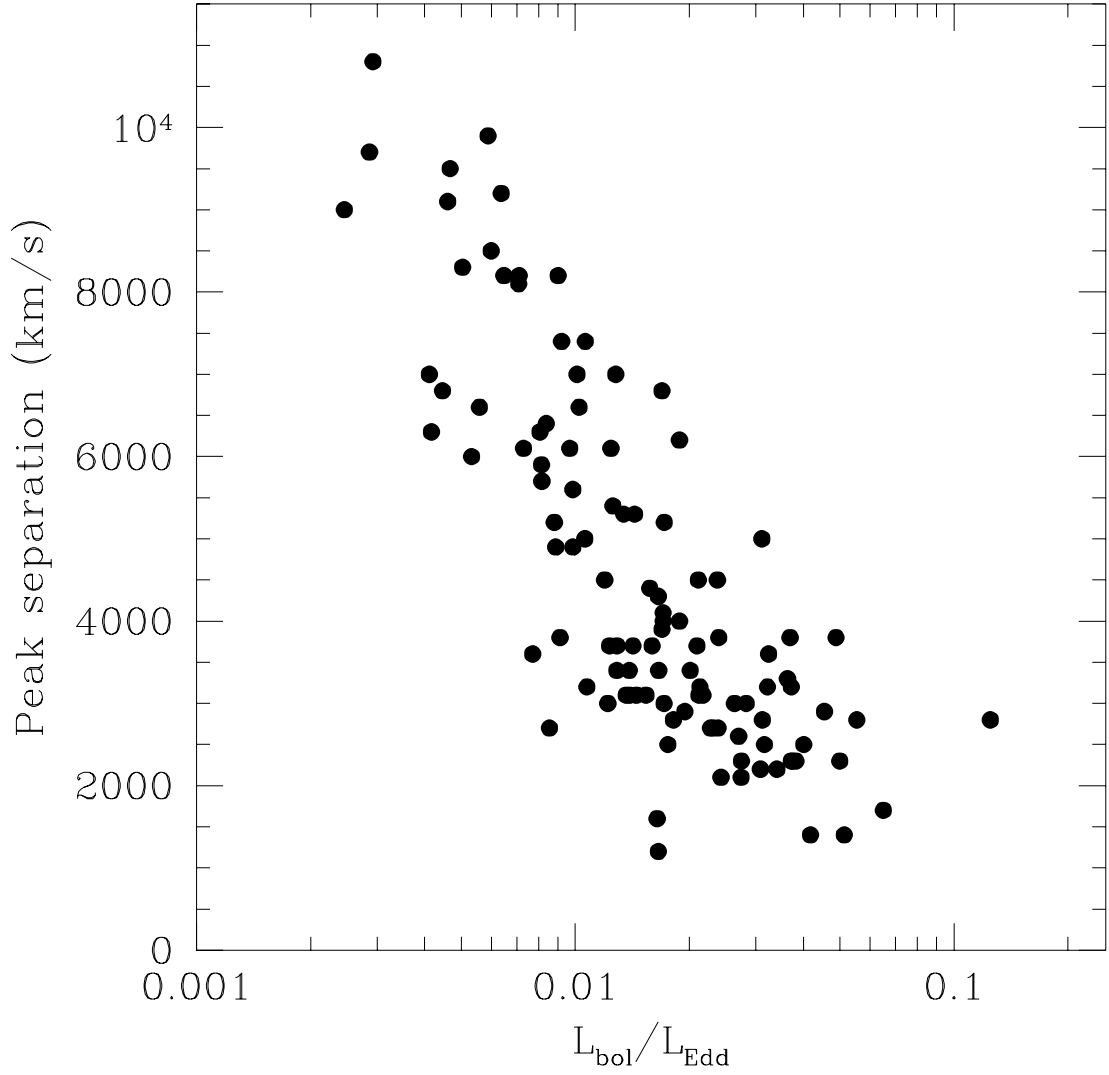}
 \caption{The relation between the peak separation in double-peaked broad line profiles
and Eddington ratio for the double-peaked SDSS AGNs. }
\end{figure}

\newpage

  \begin{table*}
\footnotesize
      \caption{Data of 116 double-peaked broad line AGNs in the SDSS sample}
         \begin{tabular}{lccccccc}
           \\ \hline
            \noalign{\smallskip}
Name & z & FWHM& $Log L_{5100\AA}$&$Log M_{BH}$& $Log~ \dot{m}$ & $\alpha_{OX}$ & Peak separation\\
     &   & (km/s)          &  (erg/s)         &  (M$_\odot$) &                      &  & (km/s)\\
\hline
         \noalign{\smallskip}

   SDSS  J0007+0053   &   0.3159   &     9800   &    44.89   &    9.408   &   -1.675      &   1.4 & 4500\\
   SDSS  J0008-1046   &   0.1986   &     9600   &     44.2   &    8.903   &   -1.866      &     ... & 3100\\
   SDSS  J0012-1022   &     0.22   &     5500   &    44.43   &    8.584   &   -1.312         &     ...& 3800 \\
   SDSS  J0043+0051   &   0.3083  &   11800   &    44.36   &    9.198   &   -1.995        &   1.5 & 7000\\
   SDSS  J0057+1446   &   0.1718   &     9900   &     44.6   &    9.213   &   -1.771     &   1.5 & 3900\\
   SDSS  J0111-0958   &   0.2064   &     5400   &    43.98   &    8.254   &    -1.43       &     ... & 2300\\
   SDSS  J0117-0111   &   0.1855   &     5600   &    43.82   &     8.17   &   -1.511       &     ... & 2200\\
   SDSS  J0132-0952   &   0.2597  &   15200   &    44.09   &    9.225   &   -2.298        &     ... & 8300 \\
   SDSS  J0134-0841   &   0.0699   &     8300   &    43.19   &    8.075   &    -2.04       &     ... & 3800\\
   SDSS  J0149-0808   &   0.2093   &     5400   &    43.73   &    8.077   &   -1.506      &   1.2 & 2800\\
   SDSS  J0212-0030   &   0.3942     &   ... &    44.63   &    ...   &     ...        &     ... & ...\\
   SDSS  J0216-0052   &   0.2778   &     9600   &    43.85   &    8.661   &   -1.969       &     ... & 3200\\
   SDSS  J0220-0728   &   0.2136   &     6100   &    44.08   &    8.427   &   -1.507       &   1.5 & 5000\\
   SDSS  J0229-0008   &   0.6091     &   ... &    44.36   &    ...   &     ...         &     ... & ...\\
   SDSS  J0232-0828   &   0.2652   &     8400   &    44.46   &    8.971   &   -1.671      &     ... &3200\\
   SDSS  J0240-0041   &   0.2466   &     9600   &    44.29   &    8.966   &   -1.839       &     ... & 3100\\
   SDSS  J0247-0714   &    0.334  &   13100   &    44.15   &    9.142   &   -2.149        &     ... & 8100\\
   SDSS  J0248-0100   &    0.184   &     6800   &    43.97   &    8.441   &   -1.636       &     ... & 2700\\
   SDSS  J0252+0043   &   0.1696   &     6400   &    44.01   &    8.423   &   -1.569       &     ... & 2600\\
   SDSS  J0259-0015   &   0.1018   &     4500   &    43.26   &    7.587   &    -1.49       &     ... & 3600\\
   SDSS  J0300-0714   &   0.3883     &   ... &    44.92   &    ...   &     ...        &     ... & ...\\
   SDSS  J0325+0008   &   0.3602  &    12000   &    44.75   &    9.485   &   -1.893      &   1.6& 7000\\
   SDSS  J0349-0626   &   0.2877   &     9900   &    44.26   &    8.977   &   -1.872       &     ... & 5300\\
   SDSS  J0739+4043   &   0.2081  &   14200   &    43.71   &    8.905   &   -2.351       &     ... & 6800\\
   SDSS  J0741+2755   &   0.3256   &     5700   &    44.31   &    8.529   &   -1.379        &     ... & 1400\\
   SDSS  J0754+4316   &   0.3475  &   10300   &    44.74   &    9.342   &   -1.765      &   1.4 & 5200\\
   SDSS  J0759+3528   &   0.2886  &   14300   &    43.62   &    8.844   &   -2.386      &     1 & 7000\\
   SDSS  J0803+2932   &   0.3277   &     9700   &    44.09   &    8.839   &   -1.906     &   1.3 & 6100\\
   SDSS  J0806+4841   &     0.37  &   12600   &    44.62   &    9.438   &   -1.974      &   1.4 & 7400\\
   SDSS  J0813+4834   &   0.2738   &     8400   &    44.23   &    8.807   &   -1.741      &     ... & 2800\\
   SDSS  J0817+3435   &    0.062  &   12600   &    43.27   &     8.49   &    -2.38       &   1.1 & 6300\\
   SDSS  J0819+4817   &   0.2228   &     9200   &    43.96   &    8.698   &   -1.901     &   1.4 & 5400\\
   SDSS  J0821+3503   &   0.2936   &     5000   &    44.19   &    8.331   &   -1.302      &     ... & 2300\\
   SDSS  J0821+4219   &    0.222  &   11200   &    43.88   &    8.818   &   -2.093     &     ... & 6300\\
   SDSS  J0821+4702   &   0.1283  &   10800   &    43.91   &    8.802   &   -2.056       &   1.1 & 5200\\
   SDSS  J0822+4553   &   0.2998   &     6200   &    44.43   &    8.683   &   -1.418       &     ... & 2300\\
   SDSS  J0824+3342   &   0.3179  &   12500   &    44.26   &    9.175   &   -2.077        &     ... & 6400\\
   SDSS  J0832+3707   &    0.092  &   15200   &    43.98   &    9.149   &   -2.331      &   1.4 & 9500\\

            \hline
         \end{tabular}
\end{table*}

\newpage
  \begin{table*}
\footnotesize
         \begin{tabular}{lccccccccc}
           \\ \hline
            \noalign{\smallskip}
Name & z & FWHM& $Log L_{5100\AA}$&$Log M_{BH}$& $Log~ \dot{m}$ & $\alpha_{OX}$ & Peak separation\\
     &   & (km/s)          &  (erg/s)         &  (M$_\odot$) &                      &  &(km/s)\\
\hline
         \noalign{\smallskip}
   SDSS  J0838+3719   &    0.211   &     9200   &    43.71   &    8.526   &   -1.975        &   1.1 & 5000\\
   SDSS  J0841+0229   &   0.3322   &     8300   &    44.06   &    8.682   &    -1.78     &   1.4 & 3400\\
   SDSS  J0845+0016   &   0.2613  &   10600   &    44.01   &    8.861   &   -2.007     &   1.2 & 5600\\
   SDSS  J0904+5536   &   0.0372     &   ... &    43.03   &    ...   &     ...         &     ... & ...\\
   SDSS  J0914+0126   &   0.1977  &   12200   &    44.32     &   9.2   &   -2.036       &     ... & 7400\\
   SDSS  J0918+5139   &   0.1855   &     4900   &    44.17     &   8.3   &    -1.29       &     ... & 1400\\
   SDSS  J0925+5317   &   0.1862   &     8000   &    44.23   &     8.77   &   -1.697       &     ... & 3400\\
   SDSS  J0935+4819   &   0.2237   &     8500   &    44.21   &    8.808   &   -1.756       &   1.4 & 2500\\
   SDSS  J0936+5331   &   0.2281   &     4800   &    44.45   &    8.481   &   -1.187       &     ... & 1700\\
   SDSS  J0938+0057   &   0.1704   &     8300   &    44.29   &    8.844   &   -1.711       &     ... & 2900 \\
   SDSS  J1000+0259   &    0.339  &   10200   &    44.43   &    9.121   &   -1.848       &     ... & 3700\\
   SDSS  J1004+4801   &   0.1986  &   13500   &    43.89   &    8.987   &   -2.253      &   1.1 & 6600\\
   SDSS  J1014+0006   &   0.1412  &   20700   &    43.94   &    9.392   &    -2.61      &     ... & 9000\\
   SDSS  J1027+6050   &   0.3314  &   16200   &    44.61   &    9.649   &   -2.196      &   1.6 & 9200\\
   SDSS  J1032+6008   &   0.2939   &     7900   &    44.26   &    8.775   &   -1.679      &   1.5 & 3700\\
   SDSS  J1041-0050   &   0.3029   &     7800   &     44.4   &    8.864   &   -1.625       &   1.4 & 4500\\
   SDSS  J1041+0232   &    0.182  &   10800   &     43.8   &    8.725   &   -2.088      &     ... & 5700\\
   SDSS  J1041+5620   &   0.2304  &   13100   &    43.91   &    8.972   &   -2.222        &     ... & 8500\\
   SDSS  J1107+0421   &   0.3269   &     9000   &    44.24   &    8.876   &   -1.797        &     ... & 3700\\
   SDSS  J1127+6750   &    0.194   &     5500   &    43.91   &     8.22   &   -1.468      &   1.3 & 2200\\
   SDSS  J1130+0058   &   0.1325   &     6900   &    44.25   &    8.656   &   -1.562        &     ... & 2100\\
   SDSS  J1130+0222   &    0.241   &     6000   &    44.26   &    8.536   &    -1.44       &     ...  & 3300\\
   SDSS  J1136+0207   &    0.239   &     5800   &     44.3   &    8.538   &   -1.397      &     ... & 2500\\
   SDSS  J1140+0546   &   0.1315  &   14400   &     43.8   &    8.977   &   -2.337         &     ... & 9100\\
   SDSS  J1143-0029   &   0.1715  &   10300   &    43.99   &    8.817   &    -1.99       &     ... & 6600\\
   SDSS  J1150-0316   &   0.1486  &   10800   &    43.92   &     8.81   &   -2.052        &     ... & 4900\\
   SDSS  J1152+6048   &   0.2703  &   10400   &    43.96   &    8.806   &   -2.007     &   1.2 & 4900\\
   SDSS  J1156+6147   &   0.2265   &     5700   &    44.15   &    8.419   &   -1.427       &     ... & 2300\\
   SDSS  J1211+6044   &    0.637     &   ... &    44.24   &    ...   &     ...   &         1.1 & ...\\
   SDSS  J1218+0200   &    0.327   &     7900   &    44.69   &    9.079   &   -1.549      &   1.5 & 3000\\
   SDSS  J1220-0132   &   0.2879   &     8800   &    44.12   &    8.774   &   -1.813      &     ... & 3100\\
   SDSS  J1238+5325   &   0.3478  &   15400   &    44.66   &    9.639   &   -2.137          &     1 & 6100\\
   SDSS  J1309+0322   &   0.2665   &     8000     &    44   &    8.609   &   -1.766      &     ... & 3000\\
   SDSS  J1324+0524   &   0.1154  &   14900   &    43.21   &    8.595   &   -2.543       &     ... & 9700\\
   SDSS  J1328-0129   &   0.1515  &   11100   &     43.8   &    8.748   &   -2.113         &     ... & 3600\\
   SDSS  J1333+0130   &   0.2171   &     8700   &    43.83   &     8.56   &   -1.891         &     ... & 3400\\
   SDSS  J1333+0418   &   0.2022   &     8000   &    43.99   &    8.603   &   -1.768     &   1.4 & 4100\\
   SDSS  J1334-0138   &   0.2917  &   16300   &    44.37   &    9.484   &   -2.274       &     ... & 6000\\
   SDSS  J1339+6139   &   0.3723  &   13400   &    44.22   &    9.209   &   -2.148      &     ... & 8200\\

            \hline
         \end{tabular}
\end{table*}

\newpage
  \begin{table*}
\footnotesize
         \begin{tabular}{lcccccccc}
           \\ \hline
            \noalign{\smallskip}
Name & z & FWHM& $Log L_{5100\AA}$&$Log M_{BH}$& $Log~ \dot{m}$ & $\alpha_{OX}$ & Peak separation\\
     &   & (km/s)          &  (erg/s)         &  (M$_\odot$) &                      & & (km/s)\\
\hline
         \noalign{\smallskip}
   SDSS  J1346+6220   &   0.1163     &   ... &    43.93   &    ...   &     ...       &   1.6 & ...\\
   SDSS  J1351+6531   &   0.2988   &     8400   &    44.13   &    8.737   &   -1.771    &   1.4 & 6800\\
   SDSS  J1400+6314   &   0.3309  &   10600   &    44.56   &    9.242   &   -1.843    &   1.5 & 5300\\
   SDSS  J1407+0235   &   0.3094  &   13900   &    44.05   &    9.124   &    -2.23       &     ... & 9900\\
   SDSS  J1414+0133   &   0.2704  &    11000   &    43.92   &    8.825   &   -2.068     &   1.2 & 2700\\
   SDSS  J1416+0219   &    0.158  &   11800   &    44.05   &    8.979   &   -2.089        &     ... & 5900\\
   SDSS  J1419+6503   &   0.1478   &     6600   &    43.92   &    8.384   &   -1.623       &     ... & 2700\\
   SDSS  J1424+5953   &   0.1348   &     5600   &    44.38   &    8.564   &   -1.342         &   1.6 & 2900\\
   SDSS  J1427+6354   &   0.1453   &     9100   &    43.85   &    8.616   &   -1.923     &   1.4 & 4500\\
   SDSS  J1434+5723   &   0.1749   &     7700   &     44.2   &    8.712   &   -1.674     &   1.3 & 3100\\
   SDSS  J1521+0337   &   0.1261   &     6600   &    43.95   &    8.403   &   -1.615       &     ... & 2100\\
   SDSS  J1540-0205   &     0.32  &   10700   &    44.98   &    9.546   &   -1.725         &     ... & 6200\\
   SDSS  J1545+5736   &   0.2681   &     7400   &    44.19   &    8.669   &   -1.643       &     ... & 2700\\
   SDSS  J1605-0109   &   0.2425   &     6700   &    43.97   &    8.432   &   -1.621       &     ... & 3800\\
   SDSS  J1635+4816   &   0.3088   &     8000   &    43.95   &    8.573   &   -1.781     &   1.5 & 1200\\
   SDSS  J1638+4335   &   0.3391   &     9000   &    44.29   &    8.915   &   -1.781       &   1.4 & 4300\\
   SDSS  J1701+3404   &   0.0945   &     6400   &     43.3   &     7.92   &   -1.784    &   1.1 & 1600\\
   SDSS  J1710+6521   &   0.3853   &     3700   &    44.64   &    8.386  &    -0.9043     &   1.9 & 2800\\
   SDSS  J1718+5933   &   0.2728  &   10400   &    43.93   &    8.788   &   -2.015     &     ...  & 6100\\
   SDSS  J1721+5344   &   0.1918   &     8200   &    43.77   &    8.466   &   -1.857     &   1.5 & 3400\\
   SDSS  J1727+6322   &   0.2175   &     8700   &    44.38   &    8.947   &   -1.725     &   1.5 & 4000\\
   SDSS  J1730+5500   &   0.2491   &     6800   &    43.88   &    8.379   &   -1.663      &     ... & 3100\\
   SDSS  J2050-0701   &   0.1686   &     5500   &    44.03     &   8.3   &   -1.433     &   1.3 & 3800\\
   SDSS  J2101-0547   &   0.1794   &     9700   &    44.08   &     8.83   &    -1.91       &   1.3 & 3700\\
   SDSS  J2113-0612   &   0.2411   &     5800   &    43.69   &    8.111   &    -1.58       &     ... & 3000\\
   SDSS  J2125-0813   &   0.6246     &   ... &    45.36   &    ...   &     ...     &   1.2 & ...\\
   SDSS  J2145+1210   &   0.1113   &     5300   &    43.49   &    7.893   &   -1.562       &     ... & 2300\\
   SDSS  J2149+1138   &   0.2393  &   17600   &    43.72   &    9.098   &   -2.534    &     ... & 10800\\
   SDSS  J2150-0010   &   0.3351   &     6200   &    44.39   &    8.655   &    -1.43     &   1.5 & 3200\\
   SDSS  J2221-0109   &   0.2878  &   14400   &     44.3   &    9.325   &   -2.188       &     ... & 8200\\
   SDSS  J2229+0008   &   0.2657   &     5200   &    44.45   &    8.549   &   -1.257       &     ... & 2800\\
   SDSS  J2233-0743   &    0.175   &     8300   &    43.81   &    8.504   &   -1.856       &     ... & 3100\\
   SDSS  J2233-0843   &   0.0582   &     4700   &    43.35   &    7.687   &   -1.501         &     ... & 2500\\
   SDSS  J2304-0841   &   0.0471   &     8600   &    43.72   &    8.472   &   -1.914        &   1.2 & 3000\\
   SDSS  J2305-0036   &   0.2687  &   11800   &     44.2   &    9.081   &   -2.046         &     ... & 8200\\
   SDSS  J2312-0116   &   0.2139   &     5800   &    43.98   &    8.315   &   -1.493       &     ... & 3200\\
   SDSS  J2327+1524   &    0.046   &     7900   &    43.84   &    8.485   &   -1.803       &     ... & 4400\\
   SDSS  J2332+1513   &   0.2146   &     8600   &     44.2   &    8.812   &   -1.768      &     ... & 4000\\
   SDSS  J2351+1552   &   0.0966   &     8700   &    43.83   &    8.562   &    -1.89     &   1.4 & 3700\\
            \hline
         \end{tabular}
\end{table*}

\newpage
  \begin{table*}
\footnotesize
      \caption{Data of 26 double-peaked broad line AGNs in the radio-loud AGN sample}
         \begin{tabular}{lccccccccc}
           \\ \hline
            \noalign{\smallskip}
Name & z & $m_V$ & $A_V$ & SF$^a$ &FWHM& $Log L_{5100\AA}$&$Log M_{BH}$& $Log~ \dot{m}$& $\alpha_{OX}$\\
     &   &  & & &(km/s)          &  (erg/s)         &  (M$_\odot$) &                       &  \\
\hline
         \noalign{\smallskip}

      3C       17 &    0.22 &      18 &   0.077 & 0.58&  11500 &   43.66 &   8.687 &  -2.182 &   1.255 \\
       4C    31.06 &   0.373 &      18 &   0.175 &  0.11&  9000 &   44.44 &   9.014 &  -1.738 &   1.238 \\
       3C       59 &   0.109 &      16 &   0.211 &  0.28&  9800 &   44.18 &   8.912 &  -1.888 &   1.248 \\
      PKS    0235+023 &   0.209 &    17.7 &   0.109  & 0.46& 11200 &   43.87 &   8.805 &  -2.099 &  0.9925 \\
     IRAS 02366--3101 &   0.063 &   14.98 &   0.218 &  0.30&  7800 &   44.13 &   8.673 &  -1.707 &   1.712 \\
      PKS    0340--37 &   0.285 &    18.6 &   0.032 &  0.19&  9800 &   43.89 &   8.708 &  -1.975 &   1.025 \\
       3C       93 &   0.357 &    19.2 &   0.804  & 0.43& 19600 &   43.98 &   9.372 &  -2.551 &    1.09 \\
       MS 0450.3--1817 &   0.059 &    17.8 &   0.144 & 0.90&  10900 &   42.07 &   7.523 &  -2.615 &   1.047 \\
  Pictor        A &   0.035 &    16.2 &   0.142  & 0.14& 18400 &    43.2 &   8.769 &  -2.731 &   1.064 \\
       B2     0742+31 &   0.462 &      16 &   0.227 &  0 &  6500 &   45.46 &   9.451 &  -1.147 &   1.346 \\
      CBS       74 &   0.092 &      16 &   0.118 &  0.17&  9200 &   44.07 &   8.776 &  -1.867 &   1.203 \\
      PKS    0857--19 &    0.36 &    19.7 &   0.685 & 0&   6500 &   43.98 &   8.415 &  -1.591 &     ... \\
      PKS   0921--213 &   0.053 &    16.5 &     0.2 & 0.65&   8300 &   43.06 &   7.983 &   -2.08 &   1.038 \\
      PKS   1020--103 &   0.197 &    16.1 &   0.153 & 0&   8700 &   44.74 &   9.201 &  -1.616 &   1.262 \\
       4C    36.18 &   0.392 &      18 &   0.056 &  0&  6800 &   44.48 &   8.798 &  -1.483 &   1.254 \\
     PKS    1151--34 &   0.258 &    17.8 &   0.284  & 0.75& 13400 &   43.73 &   8.864 &  -2.297 &   ... \\
      TXS    1156+213 &   0.349 &    17.5 &   0.088 &  0.43&  7600 &   44.36 &   8.812 &  -1.615 &   ... \\
      CSO      643 &   0.276 &    16.7 &   0.042 & 0.28&   9000 &   44.58 &   9.116 &  -1.694 &   1.136 \\
       3C      303 &   0.141 &    17.3 &   0.063 &  0.73&  6800 &   43.39 &   8.039 &  -1.808 &   1.138 \\
       3C      332 &   0.151 &      16 &   0.079  &  0.85& 23200 &   43.72 &   9.334 &  -2.776 &   1.269 \\
      Arp     102B &   0.024 &    14.8 &    0.08  & 0.90&  16000 &   42.48 &   8.141 &  -2.826 &   1.286 \\
      PKS    1739+18C &   0.186 &    17.5 &   0.206 & 0.11&  13600 &   44.11 &   9.144 &  -2.195 &   1.033 \\
       3C      382 &   0.059 &    15.4 &   0.231  & 0.31& 11800 &   44.04 &   8.969 &  -2.094 &  0.9985 \\
       3C    390.3 &   0.057 &    15.4 &   0.237  & 0.31& 11900 &   43.87 &   8.864 &  -2.149 &   1.183 \\
      PKS    1914--45 &   0.368 &    16.8 &   0.266 &  0.10&  9800 &   44.95 &   9.446 &  -1.659 &   1.595 \\
      PKS    2300--18 &   0.129 &    17.8 &   0.108 &  0.14&  8700 &   43.64 &   8.427 &  -1.948 &  0.9648 \\
            \hline
 \noalign{\smallskip}
         \end{tabular}    
$^a$~Starlight fraction values were taken from Eracleous \& Halpern (1994, 2003) 
except that a value of 0.9 instead of 1 was adopted for MS 0450.3-1817 and Arp 102B.
\end{table*}

\end{document}